\documentstyle[12pt]{article}
\newcommand\btd{\raise 2pt \hbox{$\hat\bigtriangledown$}\hskip 1.5pt}
\newcommand\bt{\raise 2pt \hbox{$\bigtriangledown$}\hskip 1.5pt}

\begin{document}

\begin{center}{\large\bf Darboux transformation for the
modified Veselov-Novikov equation}
\end{center}
\begin{center}
 Delong Yu${}^1$, Q. P. Liu ${}^2$ and  Shikun Wang${}^3$
\end{center}
\begin{center}
${}^1$ Beijing University of Chemical Technology,\\
Beijing 100029 China,\\
${}^2$ China University of Mining and Technology,\\
Beijing 100083 China.\\
${}^3$ Institute of Applied Mathematics, \\
Academy of Mathematics and System Science,\\
Chinese Academy of Science\\
P.O.Box 2734, Beijing, 100080 China, \\
{\tt wsk@amath5.amt.ac.cn}
\end{center}
\par\

\begin{abstract}
 A  Darboux transformation is constructed for the modified
Veselov-Novikov equation.
\end {abstract}

\newpage

\section{Introduction}

  Solitons and geometry are closely connected. Many
soliton equations or
integrable systems have their origins in classical differential
geometry. The best-known example, probably the first one,
is  the celebrated sine-Gordon equation, which was used to describe
surfaces with constant negative Gaussian curvature.
Another example is the
binormal flow of a curve in $R^3$. It essentially
  appeared in the study of
vortex filaments in the paper of da
  Rios \cite{da rios}. Much later,
Hasimoto \cite{ha} showed the equivalence of this
 system with the non-linear
Schr\"odinger equation. For more references on the interrelation between
geometry and integrable systems, we refer to the recent books  \cite{geo}
\cite{geo1} and the references there.

Recently,
a class of integrable  deformations of surfaces immersed in $R^3$ is defined
by using of
 generalized Weierstrass representation in \cite{k2}.
  The mean
observation of that paper is
 that the operator from the linear problem of
generalized Weierstrass
 representation
 coincides with the operator $L^{mNV}$ to which the
modified Veselov-Novikov (mVN) hierarchy is attached. Thus, the
geometrical significance of the mVN equation is established and it
is important to construct the explicit solutions for this equation.
It is well known that one of the powerful techniques leading to
explicit solutions for an integrable equation is  Darboux
transformation (DT) \cite{1}. In this paper we construct a binary DT for the
mVN
equation.

This paper is organized as follow: In section II  we will give a
brief review of the mVN equation. In section III, the derivation
of the Darboux transformation  for the mVN equation will be given and
as an example, we generate the solutions by the above Darboux
transformation for the simplest case: $U(x,y,t)=0$. We present our
conclusions and discussions in section IV.

\section{mVN equation and its Lax representation}
The mVN equation is a natural 2-dimensional generalization of MKdV
equation. The MKdV equation reads as
\begin{equation}
\label{mkdv}
       U_{t}=\frac{1}{4}U_{xxx}+6U^{2}U_{x},
\end{equation}
while the mVN equation is \cite{T}
\begin{equation}
\label{mVN}
       U_{t}=(U_{zzz}+3U_{z}V+\frac{3}{2}UV_{z})+
(U_{\bar{z}\bar{z}\bar{z}}+3U_{\bar{z}}{\bar V}+\frac{3}{2}U{\bar V}_{\bar
z}),  \quad V_{\bar{z}}=(U^{2})_{z}.
\end{equation}
It is known that the mVN equation is represented by a Manakov's triad, namely it
has the following operator formalism
\begin{equation}
L_{t}+[L,A]-BL=0,
\label{oper}
\end{equation}
where
\begin{eqnarray*}
&& L= \left(\begin{array}{cc}
  \partial&-U\\
  U&\bar{\partial}\\
  \end{array} \right), \\
&& A= \partial^{3}+\bar{ \partial}^{3}+3\left(\begin{array}{cc}
 0&-U_{z}\\
 0&V\\
  \end{array} \right)\partial
+3\left(\begin{array}{cc}
\bar{V}_{\bar{z}}&0\\
 U_{\bar{z}}&0
  \end{array} \right)\bar{\partial}
+\frac{3}{2}\left(\begin{array}{cc}
\bar{V}_{\bar{z}}&2UV\\
-2 U\bar{V}&V_{z}
  \end{array} \right), \\
&& B= 3\left(\begin{array}{cc}
 0&U_{z}\\
-U_{z}&0
  \end{array} \right)\partial+3\left(\begin{array}{cc}
0&U_{\bar{z}}\\
-U_{\bar{z}}&0
  \end{array} \right)\bar{\partial} \\
  &&\quad \quad +3\left(\begin{array}{cc}
0&U_{\bar{z}\bar{z}}+U(\bar{V}-V)\\
-U_{zz}-UV+U\bar{V}&0\\
  \end{array} \right),\\
&&\partial=\frac{1}{2}(\partial_{x}-i\partial_{y}),
\qquad \bar{\partial}=\frac{1}{2}(\partial_{x}+i\partial_{y}).
\end{eqnarray*}
In \cite{bog}, it is shown  that the system (\ref{mVN}) is related to the
Veselov-Novikov equation in the similar manner as MKdV related to KdV
system.

\par

{\it Remarks:} 1) if the function $U$ depends only on one space variable $x$,
the mVN equation (\ref{mVN}) reduces to the  MKdV equation
(\ref{mkdv}).
2) the field variable $U$ in the equation (\ref{mVN}) is assumed  as a real
valued function.

\par

Since the mVN equation possesses the operator representation (\ref{oper}),
it deforms the kernel of the operator $L$ via the
equation
\begin{eqnarray}
\label{L,A}
\left\{\begin{array}{ll}
L\Psi=0 &\\
\Psi_{t}=A\Psi &
\end{array} \right.
\end{eqnarray}
where
\begin{equation}
\Psi={\psi_{1} \choose \psi_{2}}
\end{equation}
is known as a wave function.

\section{Darboux transformation  of mVN equation}
To construct a DT for the mVN equation, we find that it is
convenient to transform the Lax pair (\ref{L,A}) into the following
form
\begin{eqnarray}
\label{lax}
\left\{\begin{array}{ll}
\Psi_{x}=J\Psi_{y}+P\Psi &\\
\Psi_{t}=-J\Psi_{yyy}-P\Psi_{yy}+Q\Psi_{y}+S\Psi &
\end{array} \right.
\end{eqnarray}
where
\[
J=i \left(\begin{array}{cc}
 1&0\\
 0&-1\\
  \end{array} \right) \equiv i\sigma_{3},  \;\;
P=\left(\begin{array}{cc}
 0 & 2U\\
 -2U & 0\\
  \end{array} \right)\equiv 2iU\sigma_{2},
\]
\[
Q=\left(\begin{array}{cc}
 -iU^{2}+3i\bar{V}&iU_{x}-2U_{y}\\
iU_{x}+2U_{y}&iU^{2}-3iV\\
  \end{array} \right),
\]
\[
S= \left(\begin{array}{cc}
(-\frac{5}{2}iU_{y}-\frac{3}{2}U_{x})U+\frac{3}{2}\bar{V}_{\bar{z}}&
-2U^{3}-2U_{yy}+\frac{1}{2}U_{xx}+\frac{i}{2}U_{xy}+3U(\bar{V}+V)\\
2U^{3}+2U_{yy}-\frac{1}{2}U_{xx}+\frac{i}{2}U_{xy}-3U(\bar{V}+V)&
\frac{5}{2}iU_{y}-\frac{3}{2}U_{x})U+\frac{3}{2}V_{z}
  \end{array} \right),
\]
and
\begin{displaymath}
\sigma_{2} = \left(\begin{array}{cc}
0&-i\\
 i&0\\
 \end{array} \right) ,\
\sigma_{3} = \left(\begin{array}{cc}
 1&0\\
 0&-1\\
 \end{array} \right)
\end{displaymath}
are Pauli matrices.

We notice that the matrices $J$, $P$, $Q$ and $S$ have the following
involution property
\begin{equation}
\left(\begin{array}{cc}
 0&-1\\
 1&0\\
  \end{array} \right)X\left(\begin{array}{cc}
 0&1\\
-1&0\\
  \end{array} \right)=\bar{X},
\end{equation}
where $X$ is one of $J, P, Q, S$.
  It is clear that if
\begin{equation}
\Psi={\psi_{1}\choose\psi_{2}}
\end{equation}
is a vector solution of (\ref{lax}) then
\begin{equation}
\Psi^{*}={-\bar{\psi}_{2}\choose\bar{\psi}_{1}}
\end{equation}
also satisfies  (\ref{lax}). Hence from a vector solution
we obtain a matrix solution of
(\ref{lax})
\begin{equation}
\label{Msol}
\left(\begin{array}{cc}\psi_{1}&-\bar{\psi}_{2}\\
\psi_{2}&\bar{\psi}_{1}
 \end{array} \right),
\end{equation}
which we also denote by $ \Psi $ for short.

We now consider the construction of a DT for
the mVN equation.

To this end, we introduce the linear system formally conjugate to
(\ref{lax})
\begin{eqnarray}
\label{con}
\left\{\begin{array}{ll}
\Phi_{x}=\Phi_{y}`J^{T}+\Phi P^{T} \label{alax} &\\
\Phi_{t}=-\Phi_{yyy}J^{T}-\Phi_{yy}P^{T}+\Phi_{y}Q^{T}+\Phi S^{T}&
\end{array} \right.
\end{eqnarray}
  It is easy to see that if $\Psi$ is a matrix solution of  (\ref{lax}) then
$\Phi=\Psi^{T}$  is a matrix solution of  (\ref{con}).

Now with a solution  $\Psi$ of the linear system  (\ref{lax}) and a
solution $\Phi$ of the linear system  (\ref{con}) we introduce a 1-form
\begin{eqnarray} \label{form}
\omega(\Phi,\Psi)=&\Phi\Psi dy+i\Phi\sigma_{3}\Psi dx
+\bigg[-i(\Phi_{yy}\sigma_{3}\Psi+\Phi\sigma_{3}\Psi_{yy}-\Phi_{y}\sigma_{3}\Psi_{y})\nonumber\\
&+\Phi_{y}P\Psi+\Phi P^{T}\Psi_{y}+\Phi\left(\begin{array}{cc}
 -iU^{2}+3i\bar{V}&iU_{x}\\
iU_{x}&iU^{2}-3iV
 \end{array} \right)\Psi\bigg]dt,
\end{eqnarray}
where $\Psi $, $\Phi $ are  matrix solutions of
(\ref{lax}), (\ref{con}) respectively. It is tedious but
otherwise straightforward to show that the 1-form defined above is
closed, that is

{\bf Lemma 1}. $d\omega(\Phi,\Psi)=0$.

{\it Proof}: By straightforward computation.

Thus, the following matrices
\begin{equation}
\label{1-form}
\hat{\Omega}(\Phi,\Psi)=\int_{M_{0}}^{M}\omega
=\int_{ (x_{0},y_{0},z_{0} ) }^{(x,y,z)}\omega,
\end{equation}
and
\begin{equation}
\label{conmatrix}
\Omega(\Phi,\Psi)=\hat{\Omega}(\Phi,\Psi)
+\left(\begin{array}{cc}
a+bi&ci\\
ci&a-bi\\
  \end{array} \right)
\end{equation}
are well defined, where $a$, $b$ and $c$ are real constants. In the
sequel, we will take  $ \Psi$ a matrix solution of  (\ref{lax}) of the
form of (\ref{Msol}).

Let $ \Psi_0 $ be any matrix solution of (\ref{lax})
of the form of (\ref{Msol}), and introduce the matrices $K$ and
$\sigma$ by
\begin{eqnarray*}
&&K\equiv\Psi_0 \Omega^{-1}(\Psi_0^{T},\Psi_0)\Psi_0^{T} =
\left(\begin{array}{cc}
k_{11}&k_{12}\\
k_{21}&k_{22}\\
  \end{array} \right), \\
&&\sigma = \Psi_{0y}\Psi_0^{-1}. \label{omega}
\end{eqnarray*}

{\bf Lemma 2}. $K$ {\it and} $\sigma$ {\it defined above have to be in the following forms}:
\begin{eqnarray*}
\label{a1}
 K = \left(\begin{array}{cc}
k_{11}&k_{12}\\
k_{12}&{\bar k}_{11}\\
  \end{array} \right),\quad
\sigma= \left(\begin{array}{cc}
\sigma_{11}&-{\bar \sigma}_{21}\\
\sigma_{21}&{\bar \sigma}_{11}
   \end{array} \right)
\end{eqnarray*}
{\it where} $k_{12}= i\lambda $ and $\lambda$ {\it is a real constant}.

{\it Proof}: From the involution property of $J$, $P$, $Q$,  $S$ and
$\Psi_0$, we find
\begin{equation}
\label{inv}
\bar{K}= \left(\begin{array}{cc}
0&-1\\
1&0\\
  \end{array} \right)K\left(\begin{array}{cc}
0&1\\
-1&0\\
  \end{array} \right),
\end{equation}
that is, $K$ also possesses the involution property. Meanwhile,
$K$ is symmetric:
\begin{equation}
\label{sym}
K=K^{T}
\end{equation}
So we have
\begin{eqnarray*}
{\bar k}_{11}=k_{22},\quad
{\bar k}_{21}=-k_{12}, \quad
k_{12}=k_{21}=i\lambda.
\end{eqnarray*}
This proves the result for $K$. Similarly, the result for $\sigma$
can be proved.

Our  DT is now conveniently formulated as

{\bf Theorem 1}. Let $\Psi$ and $\Psi_0$ be the solutions of the
linear system (\ref{lax}). Let $\Omega(\Psi_0^T,\Psi_0)$ and
$\Omega(\Psi_0^T,\Psi)$ be given by (\ref{omega})
with $\Phi=\Psi_0^T$, $\Psi=\Psi_0$ and $\Phi=\Psi_0^T$ and $\Psi=\Psi$, respectively.
Then if $\Omega^{-1}(\Psi_0^T,\Psi_0)$ is invertible, the new matrix of wave functions
defined by
\begin{eqnarray}
 \tilde{\Psi}=\Psi-\Psi_0
\Omega^{-1}(\Psi_0^{T},\Psi_0)\Omega(\Psi_0^{T},\Psi)
\end{eqnarray}
satisfies  (\ref{lax}) with the potential U,V replaced by
\begin{eqnarray}
\tilde{U}&=&U-\lambda=U+ik_{12},\\
\tilde{V}&=&V+2iUk_{12}+\bar{k}_{11}^{2}-2(\sigma_{21}k_{21}+\bar{\sigma}_{11}
\bar{k}_{11}).
\end{eqnarray}

{\it Proof}:  It is quite easy to verify that the transformed quantities
do fulfill the first equation of  (\ref{lax}). However, the
verification of  the second equation of  (\ref{lax}) is too
complex to do by hand. We did check  the validity
 by means of MAPLE.

 Thus, we establish a DT for the mVN
equation. It is easily seen that this DT is a
binary Darboux transformation.

As an example, we generate the solutions by the above Darboux transformation
for the simplest case: $U(x,y,t)=0$. Then $V(x,y,t)=0$. The
linear system (\ref{lax}) in this case is:
\begin{equation}
\Psi_{x}=J\Psi_{y}, \;\;\;\;\;
\Psi_{t}=-J\Psi_{yyy}.
\end{equation}
We take
\begin{eqnarray}
&&\Psi(x,y,t) = {0 \choose 0}, \\
&&\Psi_{0}(x,y,t) = {\psi_{1}\choose\psi_{2}}={e^{\alpha x
-i\alpha y +\alpha^3 t }\choose e^{\beta x +i\beta y +\beta^3 t}},
\end{eqnarray}
where $\alpha,\beta $ are real constants. The   matrix solution is found to
be:

\begin{equation}
\Psi_{0}(x,y,t)=\left(\begin{array}{cc}\psi_{1}&-\bar{\psi}_{2}\\
\psi_{2}&\bar{\psi}_{1}
 \end{array} \right)
 = \left(\begin{array}{cc}e^{\alpha x
-i\alpha y +\alpha^3 t }&-e^{\beta x -i\beta y +\beta^3 t}\\
e^{\beta x +i\beta y +\beta^3 t}&e^{\alpha x +i\alpha y +\alpha^3
t }
 \end{array} \right),
\end{equation}
we obtain

\begin{eqnarray*}
\tilde{U} = \frac{1}{\det\Omega}\bigg[(e^{2(\alpha x+\alpha^3
t)}-e^{2(\beta x +\beta^3
t)})(\frac{2}{\alpha+\beta}-\frac{2}{\alpha+\beta}\cos(\alpha+\beta)y
e^{(\alpha+\beta)x
 +(\alpha^3+\beta^3)t}+c) \\
-(\frac{1}{\alpha}-\frac{1}{\beta}-\frac{1}{\alpha}e^{2(\alpha
x+\alpha^3 t)}\cos 2\alpha y+\frac{1}{\beta}e^{2(\beta x +\beta^3
t)}\cos 2\beta y-2b)\cos(\beta-\alpha)y \\
-(\frac{1}{\alpha}e^{2(\alpha x+\alpha^3 t)}\sin 2\alpha
y+\frac{1}{\beta}e^{2(\beta x +\beta^3 t)}\sin 2\beta
y+2a)\sin(\beta-\alpha)y\bigg]
\end{eqnarray*}
with
\begin{eqnarray*}
\det\Omega &= &\det\Omega(\Psi_{0}^T,\Psi_{0})
=\bigg[\frac{2}{\alpha+\beta}-\frac{2}{\alpha+\beta}\cos(\alpha+\beta)y
e^{(\alpha+\beta)x
 +(\alpha^3+\beta^3)t}+c\bigg]^2 \\
&&+\bigg[-\frac{1}{2\alpha}+\frac{1}{2\beta}+\frac{1}{2\alpha}e^{2(\alpha
x+\alpha^3 t)}\cos 2\alpha y-\frac{1}{2\beta}e^{2(\beta x
+\beta^3 t)}\cos 2\beta y+b\bigg]^2 \\
&&+\bigg[\frac{1}{2\alpha}e^{2(\alpha x
 +\alpha^3 t)}\sin 2\alpha y+\frac{1}{2\beta}e^{2(\beta x
+\beta^3 t)}\sin 2\beta y+a\bigg]^2
\end{eqnarray*}
is a family solution of mVN equation involving three parameters
$a,b,c $.
\section{Conclusions}
In this paper we present a binary Darboux transformation for the
mVN equation. We also calculate the solutions
of the mVN equation using our DT by dressing the zero background.

Keeping in mind the geometrical background of the mVN equation,  it
will be interesting to construct solutions based on more sophisticated seeds
and study their geometrical implications.
This may be considered in a separate work.

\par

{\bf Acknowledgements}. This work is supported in part by National
Natural Science Foundation of China under grant number 19971094 and
Scientific Foundation of Chinese Academy of Sciences. We should like to
thank Dr. Y.K. Lau for many useful discussions.

\end{document}